\documentclass[aps,pra,floatfix,twocolumn,superscriptaddress]{revtex4-2}
\usepackage{graphicx}
\usepackage[english]{babel}
\usepackage{amsmath}
\usepackage{amssymb}
\usepackage{tensor}

\newcommand{\be}{\begin{eqnarray}}
\newcommand{\ee}{\end{eqnarray}}
\newcommand{\p}{\partial}

\begin{document}

\title{Winding real and order-parameter spaces via lump solitons of spinor BEC on sphere}

\author{Yan He}
\affiliation{College of Physics, Sichuan University, Chengdu, Sichuan 610064, China}
\email{heyan$_$ctp@scu.edu.cn}

\author{Chih-Chun Chien}
\affiliation{Department of Physics, University of California, Merced, CA 95343, USA.}
\email{cchien5@ucmerced.edu}

\begin{abstract}
The three condensate wavefunctions of a spinor BEC on a spherical shell can map the real space to the order-parameter space that also has a spherical geometry, giving rise to  topological excitations called lump solitons. The homotopy of the mapping endows the lump solitons with quantized winding numbers counting the wrapping between the two spaces. We present several lump-soliton solutions to the nonlinear coupled equations minimizing the energy functional. The energies of the lump solitons with different winding numbers indicate coexistence of lumps with different winding numbers and a lack of advantage to break a higher-winding lump soliton into multiple lower-winding ones. Possible implications are discussed since the predictions are testable in cold-atom experiments.   
\end{abstract}

\maketitle

\section{Introduction}
The theory of wrapping one object around another, known as homotopy in topology~\cite{NashBook,Nakahara}, has played an important role in characterizing physical phenomena. Topological excitations~\cite{Nakahara,Chaikin03}, such as vortices in superconductors and superfluids or defects in liquid crystals, have been characterized according to the winding number by monitoring the phase change of the order parameter around a loop enclosing the excitation. More recently, homotopy has been used to classify topological matter by counting the winding in a target space as the Brillouin zone is traversed, leading to a periodic table of topological systems~\cite{Chiu2016,Stanescu_book}. When applied to spinor Bose-Einstein condensates (BECs), homotopy helps elucidate the topological mechanisms behind exotic excitations~\cite{Ueda14}.

Here we characterize a class of solitons in a spinor BEC on a spherical shell with a topological winding number coming from a mapping that wraps the order-parameter space around the real 2D space. Historically, similar solitons, call lumps, have been studied on a plane with specific boundary condition at infinity to make the space topologically equivalent to a sphere~\cite{Manton}. Placing the lump soliton on a spherical shell will naturally realize the setup and makes the topological meaning of the solitons more transparent. Previous theoretical works~\cite{Speight95,McGlade06} considered lump solitons in a variant of the nonlinear sigma model, called the $CP^1$ model, in spherical geometry. However, those works impose rigid constraints on the order parameters. Here we will relax the constraints and propose a connection between topological lump solitons and spinor BEC. 

In cold atoms, various solitons have been realized, such as bright solitons~\cite{Khaykovich02,Eiermann04,Marchant13,Medley14}, soliton trains \cite{Strecker02}, soliton production by the Kibble-Zurek mechanism~\cite{Lamporesi13}, topological solitons in the Su-Schrieffer-Heeger model~\cite{Meier16}, soliton dynamics~\cite{Aycock17}, and ring dark solitons~\cite{Hung23} in single-component BEC, as well as solitons in two-component Bose gases~\cite{BH21} and three-component solitons in spinor BEC in elongated harmonic traps~\cite{Bersano18}. There have been theoretical studies of solitons in dipolar gases~\cite{Pedri05}, spinor BEC~\cite{Li05}, or spin-orbit coupled BEC~\cite{Pu15}, soliton formation across a phase transition \cite{Damski10}, Rydberg-induced solitons~\cite{Maucher11}, solitons in Fermi superfluid \cite{Scott11,Dutta17}, soliton engineering~\cite{Kengne21}, and topological pumping of solitons~\cite{Mostaan22}.

On the other hand, as reviewed in Ref.~\cite{Lundblad22}, spherical-shell potentials for ultracold atoms can be realized by either coupling radio-frequency excitation to multi-component atoms in harmonic potentials to create a bubble trap or producing a phase-separation structure of multi-species atoms in a harmonic trap with the core occupied by one species while the spherical shell occupied by the desired species. We envision a spinor BEC on a spherical shell will be realized and analyze its topological solitons by finding the minimal-energy solutions with quantized winding numbers. While atomic spinor BECs have been realized and analyzed extensively~\cite{Kawaguchi12,SK13}, spherical bubble traps of BEC in microgravity have recently been achieved~\cite{Carollo22}.

In the limit of infinite contact interactions, analytic solutions of the lumps with quantized winding numbers can be constructed explicitly, which elucidate the wrapping between the real space and order-parameter space via the vector of the spinor BEC wavefunctions. When the interactions are finite, the lump solitons are solutions to a set of coupled nonlinear equations. In the strong interaction regime, we still find an analytic solution using the spherical harmonics. In the intermediate and weak interaction regimes, we found another set of lumps via numerical solutions. While the lumps carry quantized winding numbers, their energies are above the uniform ground state. Our calculations suggest higher-winding lumps are energetically stable against decomposition into lower-winding lumps, which is in contrast to quantum vortices in superconductors or superfluids~\cite{FetterBook,Pethick_book}. We mention there are other means for generating topological properties via manipulations in real space, such as inducing band topology by modulating real-space patterns~\cite{PhysRevLett.109.215302,PhysRevA.97.023618,Song20}.

The rest of the paper is organized as follows. Sec.~\ref{sec:Theory} sets up a mean-field description of $F=1$ spinor BEC and outline the theory of lump solitons, which is characterized by the quantized winding number. Sec.~\ref{sec:Infinite_c0} shows analytic solutions of lump solitons in the infinite coupling-constant limit and explain the topological concept behind lump solitons. Sec.~\ref{sec:Finite_c0} shows analytic and numerical solutions to the nonlinear equation from the minimization of the energy functional. Multiple lump solitons are presented, and their excitation energies are evaluated. Sec.~\ref{sec:Implication} discusses the assumptions and approximations behind our treatment of the lump solitons and elucidates the subtle relation between lump solitons and skyrmions in spinor BEC. Sec.~\ref{sec:Conclusion} concludes our work. The Appendix gives some details of lump solitons in planar geometry.

\section{Theoretical framework}\label{sec:Theory}
According to Refs.~\cite{Kawaguchi12,SK13}, the mean-field energy functional of a three-dimensional $F=1$ spinor BEC is
\be\label{eq-F1}
\tilde{E}&=&\int d^3\tilde{x} \left\{\sum_{m=-1}^{1}\phi_m^*\left(-\frac{\hbar^2\nabla^2}{2M}-\mu_m-pm+qm^2\right)\phi_m \right. \nonumber \\
& &\left. +\frac{\tilde{c}_0}{2}n^2+\frac{c_1}{2}|\mathbf{F}|^2 \right\}.
\ee
Here $p$ and $q$ denote the linear and quadratic Zeeman splittings, $n=\sum_{m=-1}^{1}|\phi_m|^2$ is the total density, and $\mathbf{F}=(F_x, F_y, F_z)$ denotes the magnetization vector. We consider a simplified case with $c_1=0$, ignore the Zeeman splitting terms in absence of magnetic field, and assume the populations of the three components are equal by introducing a uniform chemical potential $\mu$ for all components. In this simplified case, the magnetization does not play a significant role. Both $\mu$ and $c_0$ are positive for a stable BEC.
The energy functional~\eqref{eq-F1} of a 3D $F=1$ spinor BEC after the simplifications  becomes
\be
\tilde{E}=\int d^3\tilde{x} \left\{\sum_{m=-1}^{1}\phi_m^*\left(-\frac{\hbar^2\nabla^2}{2M}-\mu\right)\phi_m +\frac{\tilde{c}_0}{2}n^2\right\}.
\ee

Next, we consider the $F=1$ spinor BEC confined on a spherical surface $\tilde{S}^2$ with radius $R$ and replace $\tilde{x}^\nu\to Rx^\nu$ with $\nu=1, 2$. The derivatives are scaled accordingly. Moreover, $\tilde{\phi}_a\to\mathcal{M}\phi_a$ and $\tilde{E}\to E\tilde{E}_0$, where $\tilde{E}_0=\hbar^2\mathcal{M}^2/(2M)$ and $\mathcal{M}^2=\mu/c_0$. 
The energy functional then becomes
\be\label{Eq:Etilde}
\tilde{E}&=&\int_{\tilde{S}^2}d^2\tilde{x} \sqrt{\tilde{g}}\Big[\frac{\hbar^2}{2M}\tilde{g}^{\mu\nu}\p_\mu\tilde{\phi}_a \p_\nu\tilde{\phi}_a-\mu\,\tilde{\phi}_a^2+ \frac{\tilde{c}_0}{2}(\tilde{\phi}_a^2)^2\Big]. \nonumber \\
\ee
In absence of external gauge field such as electromagnetic field or rotation, the charge-neutral bosonic system may be described by real-valued fields~\cite{Peskin_book}, so we simplify the condensate wavefunctions $\phi_{m}$ with $m=-1,0,1$ as real scalar fields and relabel them as $\tilde{\phi}_a$ with $a=1,2,3$. The repeated indices imply summation. The coordinates $\tilde{x}^{\nu}$ with $\nu=1,2$ parameterize the spherical surface.
$\tilde{g}_{\mu\nu}$ is the metric of the sphere, and $\tilde{g}=\det(\tilde{g}_{\mu\nu})$. The integration is over the whole sphere $S^2$ in real space. In the superfluid phase, both $\mu$ and $c_0$ are assumed to be positive constants. 

After adding a constant term to the above energy functional, the energy functional can be cast into the following form
\be
\tilde{E}=\int_{\tilde{S}^2}d^2\tilde{x} \sqrt{\tilde{g}}\Big[\frac{\hbar^2}{2M}\tilde{g}^{\mu\nu}\p_\mu\tilde{\phi}_a \p_\nu\tilde{\phi}_a+\frac{\tilde{c}_0}{2}(\tilde{\phi}_a\tilde{\phi}_a-\mathcal{M}^2)^2\Big].
\label{eq-en}
\ee
The repeated indices imply summation.
Here we define $\mathcal{M}^2=\mu/c_0$ for convenience, which should not be confused with the mass $M$ of the atoms. 
To rewrite the energy functional in a dimensionless form, we note that the dimension of $2M E/\hbar^2$ is $[\text{Length}]^{-2}$. Thus, $2M \tilde{c}_0/\hbar^2$ is dimensionless and $\mathcal{M}$ has dimension $[\text{Length}]^{-1}$. Following the scaling with $\tilde{g}_{\mu\nu}\rightarrow Rg_{\mu\nu}$, we obtain the dimensionless energy functional
\be
E=\int_{S^2}d^2x \sqrt{g}\Big[g^{\mu\nu}\p_\mu\phi_a \p_\nu\phi_a+\frac{c_0}{2}(\phi_a^2-1)^2\Big].
\label{eq-en-1}
\ee
Here we define $c_0=2M \tilde{c}_0(R\mathcal{M})^2/\hbar^2$. The integral is over the unit sphere $S^2$ with the metric $g_{\mu\nu}$ and $g=\det(g_{\mu\nu})$. 

The energy functional is similar to the $O(3)$ non-linear sigma model \cite{Polyakov,Rajaraman}, which can support topological solitons called ``lumps'', as discussed in  Ref.~\cite{Manton}.
For the spinor BEC on a sphere, the vector field formed by $\phi^a(x)$ defines a map from the spatial two-sphere parameterized by $x=(x^1, x^2)$ to the sphere of $\phi^a\phi^a=1$ in the order-parameter space. The topology of this mapping is given by the homotopy group $\pi_2(S^2)=\mathbb{Z}$ \cite{NashBook,Nakahara}, which can be characterized by the integer winding number
\be\label{eq-WN}
N=\frac1{8\pi}\int_{S^2}d^2x \epsilon_{\mu\nu}\epsilon_{abc}\phi_a \p_\mu\phi_b \p_\nu\phi_c.
\ee
Here $\epsilon_{\mu\nu}$ and $\epsilon_{abc}$ are the Levi-Civita symbols of the real-space coordinates and order-parameter indices, respectively.
We remark that the sign of $N$ reflects the sign convention of the condensate wavefunctions. Since Eq.~\eqref{eq-en-1} is invariant with respect to $\phi_a\to -\phi_a$, $\pm N$ comes in pairs for each solution.

\section{Spherical lump soliton in the infinite $c_0$ limit}\label{sec:Infinite_c0}
To appreciate the topological properties of the lump soliton, we first consider the limit of $c_0\to\infty$, where exact solutions exist and can be visualized. 
The condition $\phi_a\phi_a=1$ must be satisfied to minimize the energy functional if $c_0\rightarrow\infty$. Thus, the above spinor BEC model becomes the genuine $O(3)$ non-linear sigma model on a spherical shell. 
In the literature~\cite{Manton}, the lumps are defined on a 2D plane $\mathbb{R}^2$, where the condition $\phi_a\to$ constant as $|x|\to\infty$ has been imposed to map the plane to an infinite sphere. A brief overview of the planar case is given in the Appendix. Here the spinor BEC is placed on a finite spherical shell.
We now look for configurations of $\phi_a(x)$ that minimize the total energy and also give rise to non-vanishing integers of the winding number $N$.

It is convenient to convert $\phi_a$ to a complex field according to the stereographic projection defined by
\be
R(x)=\frac{\phi_1+i\phi_2}{1+\phi_3},\quad
R^*(x)=\frac{\phi_1-i\phi_2}{1+\phi_3}.
\ee
The spherical coordinates with $x_1=\theta$ and $x_2=\varphi$ correspond to the metric on a unit sphere given by $g_{\mu\nu}=diag(h_1^2, h_2^2)$ with $h_1=1$ and $h_2=\sin\theta$.
We will focus on the case when the Bogomolny bound \cite{Belavin} is saturated in this section. 
Following the stereographic projection, the energy functional ~\eqref{eq-en-1} of the spinor BEC becomes
\be
E&=&
\int_{S^2}d^2x \sqrt{g}\frac{4g^{\mu\nu}\p_\mu R\p_\nu R^*}{(1+|R|^2)^2} \nonumber \\
&=&\int_{S^2}d^2x \frac{4}{(1+|R|^2)^2}\Big(\frac{h_2}{h_1}|\p_1 R|^2+\frac{h_1}{h_2}|\p_2 R|^2\Big).
\label{eq-F}
\ee
Similarly, the winding number can be written as
\be
N=\frac{1}{2\pi}\int_{S^2}d^2x\frac{i(\p_1 R\p_2 R^*-\p_2 R\p_1 R^*)}{(1+|R|^2)^2}.
\label{eq-N}
\ee
We remark that the definition of the winding number does not depend on the metric, which reflects its topological origin. 
From Eqs.~(\ref{eq-F}) and (\ref{eq-N}), we find the so-called Bogomolny bound \cite{Belavin} 
\be\label{eq-BB}
E\ge 8\pi N.
\ee

Since
\be
\Big(\frac{h_2}{h_1}|\p_1 R|^2-\frac{h_1}{h_2}|\p_2 R|^2\Big)^2>0,
\ee
we obtain the following inequality
\be
\Big(\frac{h_2}{h_1}|\p_1 R|^2+\frac{h_1}{h_2}|\p_2 R|^2\Big)&\ge& 2|\p_1 R||\p_2 R| \nonumber \\
&\ge& 2\text{Im}\Big(\p_1 R\,\p_2 R^*\Big). 
\ee
Along with the saturation of the Bogomolny bound~\eqref{eq-BB}, the conditions of equality are reduced to the following equations
\be
h_2^2|\p_1 R|^2=h_1^2|\p_2 R|^2,~
\p_1 R\p_2 R^*+\p_2 R\p_1 R^*=0,
\ee
which can be simplified to 
\be\label{eq-Req}
h_2\p_1 R=-i h_1\p_2 R,\quad
h_2\p_1 R^*=i h_1\p_2 R^*.
\label{eq-CR}
\ee
Since the equations are complex conjugates, we only focus on the first equation.

\begin{figure}
\centering
\includegraphics[width=0.45\columnwidth]{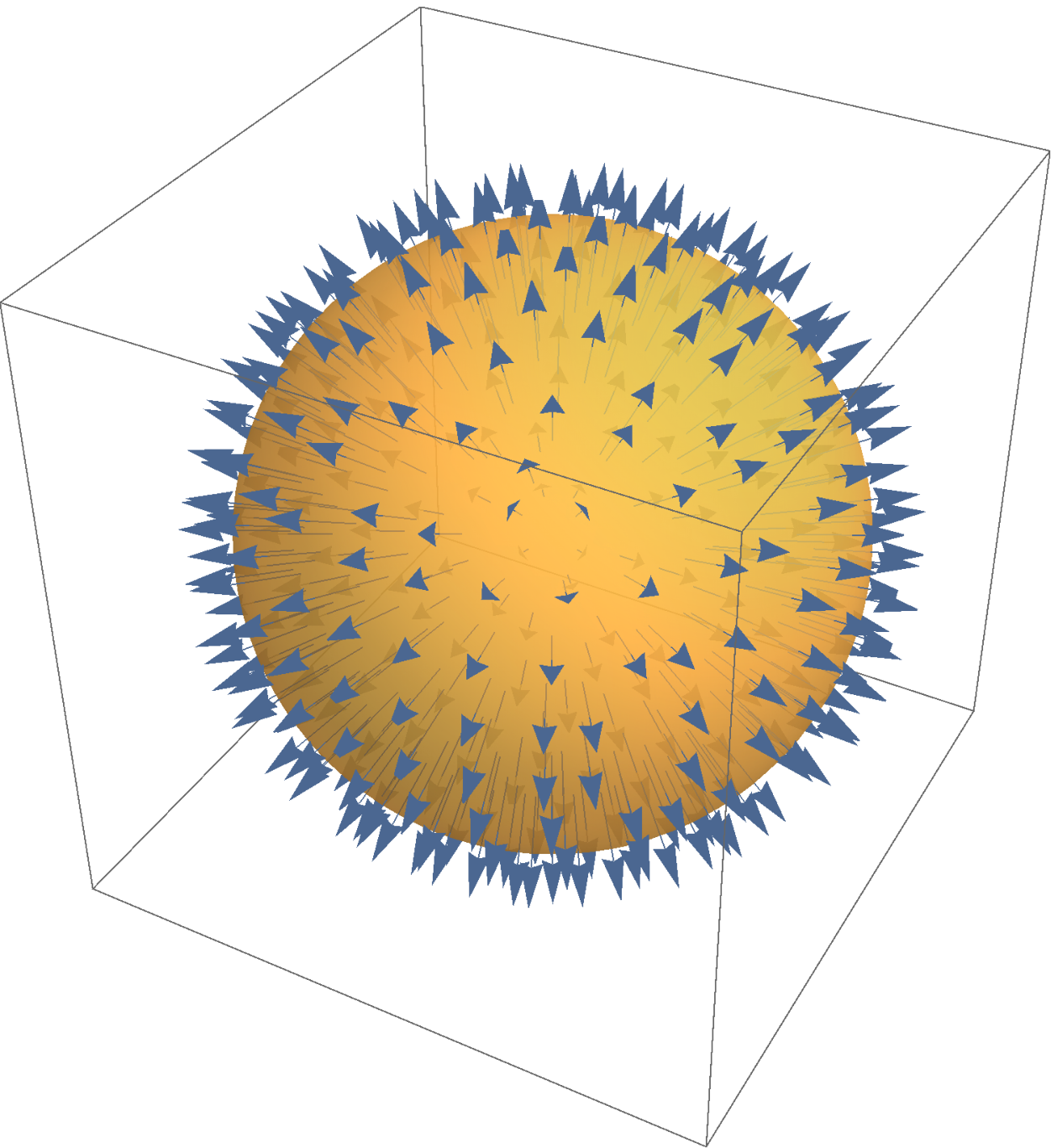}
\includegraphics[width=0.45\columnwidth]{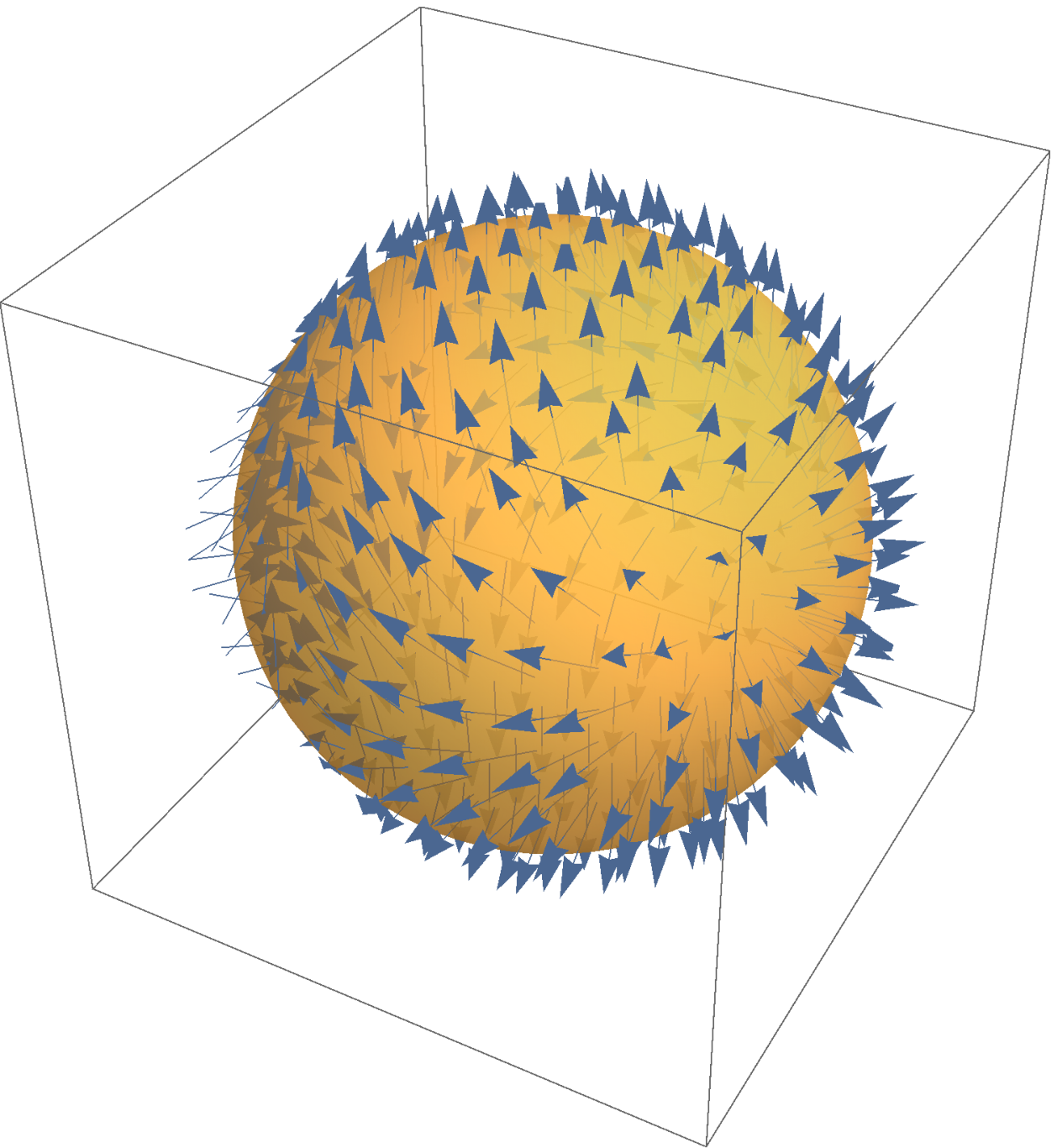}
\caption{Vector fields of the condensate wavefunctions $\phi_a$ for the $N=1$ lump soliton~\eqref{eq:N1Inf} (left) and the $N=2$ lump soliton~\eqref{eq:N2Inf} (right) with infinite $c_0$ on a unit sphere. The vectors wrap the order parameter space once (left) and twice (right).}
\label{fig-sphere}
\end{figure}

From Eq.~\eqref{eq-CR} and $u=\tan\frac{\theta}{2}e^{i\varphi}$, it follows 
$\Big(\sin\theta\frac{\p}{\p\theta}+i\frac{\p}{\p\varphi}\Big)u=0$.
Thus, the generic lump solution on a unit sphere can be taken as
\be
R(u)=\frac{p(u)}{q(u)},
\ee
where $p(u)$ and $q(u)$ are polynomials of $u$ with no common factors. Clearly, $R(u)$ satisfies Eq.~(\ref{eq-CR}). In terms of $R(u)$, we find 
$\phi_1=\frac{2\text{Re}(R)}{1+|R|^2}$, $\phi_2=\frac{2\text{Im}(R)}{1+|R|^2}$, and $\phi_3=\frac{1-|R|^2}{1+|R|^2}$.
As an example, we consider a general solution describing a single lump centered at $(\theta_0,\varphi_0)$ with winding number $N=1$ given by
$R(u)=C(u-u_0)$.
Here $C$ is a complex constant and $u_0=\tan(\theta_0/2)e^{i\varphi_0}$. For simplicity, we can assume $C=1$ and $u_0=0$, so $R(u)=u$ that translates to the configuration of the $\phi_a$ fields as
\be\label{eq:N1Inf}
\phi_1=\sin\theta\cos\varphi,\quad \phi_2=\sin\theta\sin\varphi,\quad \phi_3=\cos\theta.
\ee
This is nothing but the identity map from the real-space $S^2$ to the order-parameter space $S^2$ defined by $\phi_a^2=1$. 
This simple soliton solution is visualized in the left panel of Figure \ref{fig-sphere}, where the vector field of $\phi_a$ on a unit sphere is represented by the arrows spreading around the whole surface. The configuration is similar to the electric field of a point charge or a hedgehog.

A $N=2$ lump solution can be constructed by setting $R(u)=u^2$, which translates to the configuration of the $\phi_a$ fields as
\be\label{eq:N2Inf}
\phi_1&=&\frac{\sin^2\theta}{1+\cos^2\theta}\cos2\varphi,\quad
\phi_2=\frac{\sin^2\theta}{1+\cos^2\theta}\sin2\varphi, \nonumber \\ \phi_3&=&\frac{2\cos\theta}{1+\cos^2\theta}.
\ee
A direct calculation verifies that the winding number is indeed $N=2$ for this solution. The configuration of the $\phi_a$ fields of this lump solution is shown in the right panel of Figure \ref{fig-sphere}, where the order-parameter space wraps twice around the real space. The energies of the $N=1$ and $N=2$ lump solitons are $E_1=8\pi$ and $E_2=16\pi$, respectively. The linear energy dependence on $N$ is different from the quadratic dependence of the vortex energy on its winding number~\cite{FetterBook,Pethick_book} and implies energy neutrality by converting a $N=2$ lump to two $N=1$ lumps in the $c_0\rightarrow\infty$ limit.

\section{Spherical lump solitons with finite $c_0$}\label{sec:Finite_c0}

In realistic cases with finite $c_0$, a variation of the energy functional (\ref{eq-en-1}) gives
\be
\delta E=2\int_{S^2}d^2x \sqrt{g}\Big[g^{\mu\nu}\p_\mu\phi_a \p_\nu\delta\phi_a
+c_0(\phi_b^2-1)\phi_a\delta\phi_a\Big].
\ee
After integrating by parts and noticing that $\delta\phi_a$ is an arbitrary variation, 
we find the following set of non-linear differential equations with $a=1,2,3$. 
\be
-\frac{1}{\sqrt{g}}\p_\mu\Big(\sqrt{g}g^{\mu\nu}\p_\nu\phi_a\Big)+c_0(\phi_b\phi_b-1)\phi_a=0.
\label{eq-fB}
\ee
The extremum condition admits a trivial solution $\phi_a=v_a$, where $(v_1, v_2, v_3)$ is a constant unit vector in 3D. For the trivial solution, the energy is zero, which indicates it is the ground state without any excitation. 
We will present several solutions of the equations with non-zero winding numbers. However, finding solutions of non-linear differential equations is in general very difficult. For convenience, we rewrite the spherical Laplacian operator $\nabla_s^2\equiv-\frac{1}{\sqrt{g}}\p_\mu\sqrt{g}g^{\mu\nu}\p_\nu$ on a unit sphere as $-\Big(\frac{1}{\sin\theta}\frac{\p}{\p\theta}\sin\theta\frac{\p}{\p\theta}+\frac{1}{\sin^2\theta}\frac{\p^2}{\p\varphi}\Big)$.
The eigenfunctions of the spherical Laplacian operator are the spherical harmonics:
$\nabla_s^2Y_{lm}(\theta,\varphi)=l(l+1)Y_{lm}(\theta,\varphi)$.

We caution that Eq.~\eqref{eq-fB} is a set of nonlinear equations, which may admit multiple solutions. In terms of $Y_{lm}(\theta,\varphi)$, we can find one solution of Eq.~(\ref{eq-fB}) inspired by the infinite-$c_0$ solution~\eqref{eq:N1Inf} as
\be
&&\phi_1=A\sin\theta\cos\varphi=\frac{A}{2}(Y_{1,1}+Y_{1,-1}),\nonumber\\
&&\phi_2=A\sin\theta\sin\varphi=\frac{A}{2i}(Y_{1,1}-Y_{1,-1}),\nonumber\\
&&\phi_3=A\cos\theta=AY_{1,0}.
\label{eq-s1}
\ee
Here $A$ is a real constant, and we ignore the normalization factors of $Y_{lm}$ at this moment. Then $\phi_a^2=A^2$, and Eq.~(\ref{eq-fB}) reduces to
$2\phi_a+c_0(A^2-1)\phi_a=0$.
Hence, we find $A=\sqrt{1-\frac2{c_0}}$, which requires that $c_0>2$. One can verify that the winding number for this solution is $N=1$, and its configuration is similar to Fig.~\ref{fig-sphere} (a).

Making use of the extremum condition \eqref{eq-fB}, the kinetic energy takes the following form
\be
E_{\text{kin}}&=&\int_{S^2}d^2x\phi_a\Big(-\p_\mu\sqrt{g}g^{\mu\nu}\p_\nu\Big)\phi_a \nonumber \\
&=&-\int_{S^2}d^2x\sqrt{g}c_0(\phi_b^2-1)\phi_a^2.
\ee
Combining the above result with the potential term, the total energy with finite $c_0$ becomes
\be
E&=&E_{\text{kin}}+\int_{S^2}d^2x\sqrt{g}\frac{c_0}{2}(\phi_b^2-1)^2\nonumber\\
&=&\int_{S^2}\sqrt{g}\Big[\frac{c_0}{2}(1-\phi_a^2)(1+\phi_b^2)\Big].
\ee
For the $N=1$ lump soliton solution from Eq.~(\ref{eq-s1}), we have $\phi_b^2=A^2<1$, and the energy is 
$E=8\pi\Big(1-\frac{1}{c_0}\Big)>0$.
Therefore, the lump soliton is an excitation with positive energy, implying that the existence of a soliton solution requires some finite energy provided to the spinor BEC. Moreover, the excitation energy approaches the Bogomolny bound when $c_0\rightarrow\infty$. We remark that searching for a $N=2$ lump solution from an analogue of Eq.~\eqref{eq:N2Inf} with finite $c_0$ is challenging because there are five $Y_{2,m}$ functions but only three $\phi_a$ components. In the following, we will use a different ansatz to find more lump solutions. 

\begin{figure}[t]
\centering
\includegraphics[width=\columnwidth]{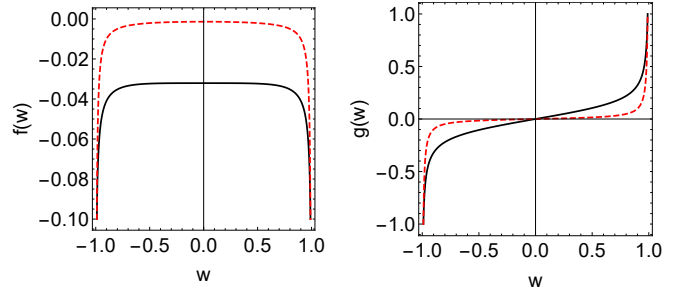}
\caption{$f$ (left panel) and $g$ (right panel) of the $N=-1$ (Black line) and $N=-2$ (Red dashed line) solutions of Eq.~(\ref{eq-phi}) on a unit sphere as functions of $w=\cos\theta$. Here $c_0=1$.}
\label{fg}
\end{figure}

We look for more solutions to the nonlinear equation~\eqref{eq-fB}.
Inspired by the $N=1$ lump solution~\eqref{eq-s1} from the spherical harmonics, we seek a more general solution of Eq.~(\ref{eq-fB}) with the following form
\be
\phi_1=f(\theta)\cos(n\varphi),\quad \phi_2=f(\theta)\sin(n\varphi),\quad \phi_3=g(\theta).
\label{eq-phi}
\ee
Here $f(\theta)$ and $g(\theta)$ are two unknown functions to be determined, and $n$ is some positive integer. By defining $w=\cos\theta$ and substituting the functional form into Eq.~(\ref{eq-fB}), we find two coupled equations:
\be
&&\frac{d}{dw}\Big[(1-w^2)\frac{df}{dw}\Big]-\frac{n^2}{1-w^2}f-c_0(f^2+g^2-1)f=0, \nonumber \\
&&\frac{d}{dw}\Big[(1-w^2)\frac{dg}{dw}\Big]-c_0(f^2+g^2-1)g=0. \label{eq-fg}
\ee
Since some coefficients of the equations are singular at $w=\pm1$, we impose the boundary condition allowed on a sphere that both $f(\pm1)$ and $g(\pm1)$ are finite. 
Inspired by the behavior of the Legendre polynomials, we test boundary conditions such as $f(-1^+)=f(1^-)=C_1$ and $g(-1^+)=-g(1^-)=C_2$ with  $C_{1,2}$ being trial constants. Different from solving linear equations, convergence of solutions to the nonlinear equations~\eqref{eq-fg} only occurs with certain values of $C_{1,2}$. We have checked that the numerical solutions obtained this way satisfy the nonlinear equation ~\eqref{eq-fg} with an error less than $10^{-5}$.

The numerical solutions of $f(w)$ and $g(w)$  with $n=1$ and $c_0=1$ are shown in Figure \ref{fg}. 
Since the norm of the vector $\phi_a$ is no longer a constant, we introduce the unit vector $\hat{\phi}_a=\phi_a/\sqrt{\phi_a^2}$, which defines a map from the real-space $S^2$ to the unit sphere of $\hat{\phi}_a$. Then we can evaluate the winding number using Eq.~\eqref{eq-WN}, which becomes
\be
N
=\frac{n}{2}\int_{-1}^1 dw\,\sqrt{1-w^2}\frac{f(g f'-f g')}{(f^2+g^2)^{3/2}}
\ee
with $f'=df/dw$. For the $n=1$ solution of Eq.~\eqref{eq-phi}, we find $N=-0.97$, which is close to $N=-1$ and indicates the result as a $N=-1$ lump soliton on a sphere. We emphasize that this is a different solution from the $N=1$ lump solution shown in Eq.~\eqref{eq-s1}. Therefore, we use $N=-1$ to denote this solution although the sign is irrelevant to the solution.

\begin{figure}[t]
\centering
\includegraphics[width=\columnwidth]{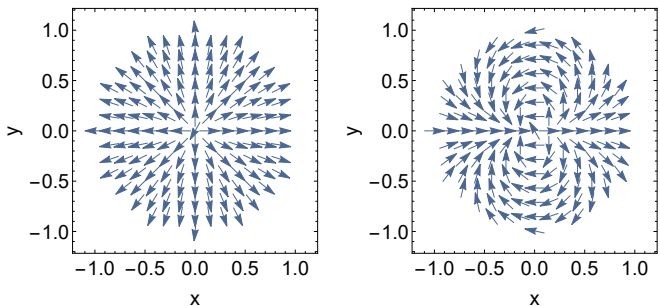}
\caption{Projections of the $\phi_a$ fields from the north hemisphere onto the $xy$ plane for the $N=-1$ (left) and $N=-2$ (right) lump soliton solutions of Eq.~\eqref{eq-phi}. All projected vectors have been normalized to unit length.}
\label{fig-vec}
\end{figure}

For a better visualization of the vector field $\phi_a$ on a sphere, we make a projected vector-field plot on a plane. To this end, we first extract the $x$ and $y$ coordinates on the north hemisphere of the real-space $S^2$ and then project the corresponding three-component $\phi_a$ vectors to the $xy$ plane. To make all vectors equally visible, we normalize the projected $\phi_a$ vectors to unit length. The $N=-1$ result is shown in the left panel of Figure \ref{fig-vec}. Taking the south hemisphere produces the same projected vector-field plot. 
If one follows a loop in real space circling the north pole once, the vectors $\hat{\phi}_a$ will make an opposite $2\pi$ rotation, which is the signature of the winding number $N=-1$.

By plugging $n=2$ in Eq.~(\ref{eq-phi}), we find a higher-winding soliton solution. The numerical results of $f(w)$ and $g(w)$ are shown by the dashed lines in Figure \ref{fg}. From the plots, one can see that the $f$ and $g$ curves are steeper around $w=\pm1$.
The numerical value of the winding number of the $n=2$ solution is $N=-1.98$, which is close to $N=-2$ and confirms the solution as a $N=-2$ lump soliton on a sphere. 
The projected vector-field plot of the $N=-2$ lump soliton is shown in the right panel of Fig.~\ref{fig-vec}. One can see the double-wrapping of the vectors around a point, reflecting the $N=-2$ winding number. When comparing Fig.~\ref{fig-vec} with finite $c_0$ to Fig.~\ref{fig-sphere} with infinite $c_0$, one can see their resemblance, despite the opposite directions of the vectors due to the opposite signs of the winding numbers.

\begin{figure}[t]
\centering
\includegraphics[width=0.8\columnwidth]{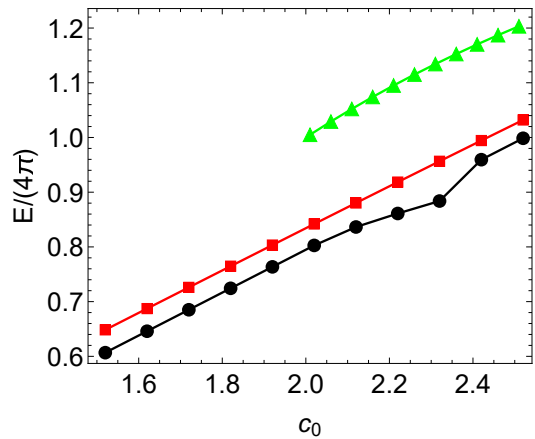}
\caption{The lump soliton energy $E=\tilde{E}/\tilde{E}_0$ on a unit sphere as a function of $c_0$ for the numerical solutions with $N=-1$ (black dots) and $N=-2$ (red squares) from Eq.~\eqref{eq-fg} and the analytical solution with $N=1$ (green triangles) from Eq.~\eqref{eq-s1}. 
}
\label{fig-E-c0}
\end{figure}

In principle, one can obtain lump solitons on a sphere with higher winding numbers by using larger values of $n$ in Eq.~\eqref{eq-phi}. However, the coefficients of Eq.~(\ref{eq-fg}) become more singular for larger values of $n$. Therefore, the higher-$n$ solutions will appear more singular around $w=\pm1$, which then lead to larger numerical errors for those high-winding solutions. 

The energies of the lump solitons are above the ground-state energy because they are topological excitations. Thus, some energy according to Eq.~\eqref{eq-en-1} is required to maintain a soliton configuration.
In Figure \ref{fig-E-c0}, we show the energies as functions of $c_0$ of the $N=-1$ and $N=-2$ lump solutions of Eq.~(\ref{eq-fg}) and also that of the $N=1$ lump solution of Eq.~(\ref{eq-s1}), which is valid only for $c_0>2$. All the excitation energies increase with $c_0$. Both $N=-1$ and $N=-2$ solutions of Eq.~(\ref{eq-fg}) have lower energies than that of the $N=1$ solution of Eq.~(\ref{eq-s1}) when $c_0 >2$.  We emphasize again that the $N=1$ lump of Eq.~(\ref{eq-s1}) is a different solution  from the $N=-1$ lump of Eq.~(\ref{eq-fg}) because the nonlinear equations admit multiple solutions.  Moreover, the $N=-2$ lump energy from Eq.~(\ref{eq-fg}) is higher than the $N=-1$ lump energy but less than twice of the $N=-1$ energy, which implies that it is not energetically favorable for a $N=-2$ lump to decay into two $N=-1$ lumps. This is again in contrast to quantum vortices with excitation energies proportional to the square of the winding number~\cite{FetterBook,Pethick_book}, which energetically favor vortices with lower winding numbers.

\section{Implications}\label{sec:Implication}
The condensate wavefunctions $\phi_a$ determines the corresponding densities, so the lump solitons may be viewed as particular density profiles of the three components. 
To induce a lump soliton, one may need to generate inhomogeneous and non-overlapping density profiles of the three components. Since the lumps are minimal-energy solutions with globally defined winding numbers, they will last longer while other non-soliton modes decay away. 
For example, one may use focused laser lights coupled to different components to induce local perturbations, similar to the methods for measuring critical velocity~\cite{Raman99,Desbuquois12,Weimer15} or exotic sounds (see Ref.~\cite{Hu22} for a review) in cold atoms. It may also be possible to use optical Feshbach resonance~\cite{Fatemi00,Theis04,Enomoto08,Bauer09,Blatt11,YanOFR13} or other means to locally tune the spin-dependent coupling $c_1$ before turning it off. 
Imaging the density profiles of different components of a spinor BEC may be performed by the method in Ref.~\cite{Bersano18}, where non-topological solitons of quasi-1D $F=1$ spinor BECs have been observed. 


We remark on extensions of the framework. While the kinetic energy is typically negligible when compared to the interaction energy in bosonic systems, the Fermi energy of fermions may remain competitive in the presence of interactions~\cite{FetterBook}. A curvature-induced BCS-BEC crossover on a spherical shell due to the competition of the kinetic and interaction energies has been discussed~\cite{He22}. Whether the lumps can survive in fermionic systems is an intriguing question. On the other hand, the homotopy $\pi_3 (S^2)=\mathbb{Z}$~\cite{Nakahara} implies a $F=1$ spinor BEC may also exhibit topological behavior if the geometry is $S^3$, which may be possible by engineering the 4D space-time.

The assumption of real-valued condensate wavefunctions allows a homotopy mapping between the real and order-parameter spaces. For spinor BEC in elongated harmonic traps, the dark-bright soliton has real-valued condensate wavefunctions~\cite{Bersano18} while other types of solitons may have complex-valued wavefunctions in the presence of magnetic field or other artificial gauge fields. If complex-valued $\phi_a$ solutions in the presence of external gauge fields are considered, the order-parameter space becomes a higher-dimensional one, which likely will render the homotopy between the real space and order-parameter space trivial. Although complex-valued soliton solutions from minimization of the free-energy may still arise, they may not be endowed with concise topological meanings like the lump solitons considered here. However, topological properties of complex-valued condensates of spinor BEC may arise in a different way, as explained below.

The vector field $\phi_a$ with $a=1,2,3$ of the wavefunction comes from the condensate wavefunction $\phi_m$ with magnetic number $m=-1,0,1$. They represents the amplitude of the densities of the three condensate components. Meanwhile, one may consider the magnetization vector $\textbf{F}$ as a vector field~\cite{Kawaguchi12}:
\begin{eqnarray}
F_x&=&\frac{1}{\sqrt{2}}[\phi_1^*\phi_0+\phi_0^*(\phi_{1}+\phi_{-1})+\phi^*_{-1}\phi_0], \nonumber \\
F_y&=&\frac{i}{\sqrt{2}}[-\phi_1^*\phi_0+\phi_0^*(\phi_{1}-\phi_{-1})+\phi^*_{-1}\phi_0], \nonumber\\
F_z&=&|\phi_{1}|^2-|\phi_{-1}|^2
\end{eqnarray}
and discuss possible topological mapping from it. As a consequence, there are two order-parameter spaces of the spinor BEC, one from the vector $(\phi_1,\phi_2,\phi_3)$ in the absence of $c_1$ and magnetic or gauge field and the other from the magnetization $\mathbf{F}$. In our setup with $c_1=0$ and no magnetic field, $\mathbf{F}$ does not affect the free energy and as a consequence, the topology is characterized by the vector field $\phi_a$.

\begin{figure}[t]
\centering
\includegraphics[width=0.8\columnwidth]{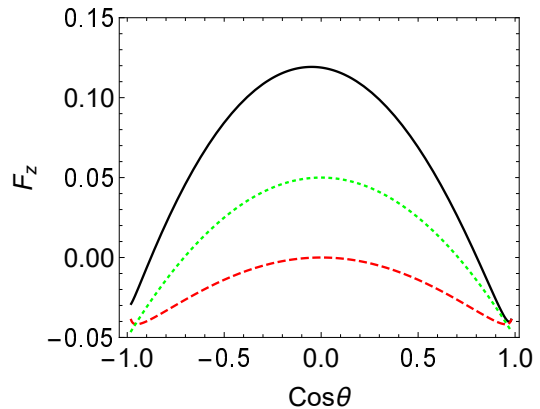}
\caption{The $z$ component magnetization $F_z$ as a function of $\cos\theta$ with $\varphi=0$ for the numerical solutions with $N=-1$ (black solid line) and $N=-2$ (red dashed line) and the analytical solution with $N=1$ (green dotted line). Here $c_0=2.3$ and $c_1=0$. }
\label{fig-Fz}
\end{figure}

We remark that the coupling constants $c_0$ and $c_1$ are related to the scattering lengths $a_0$ and $a_2$ via~\cite{SK13}
$c_0=\frac{4\pi\hbar^2}{M} \frac{a_0+2a_2}{3}$ and $c_1=\frac{4\pi\hbar^2}{M} \frac{a_2-a_0}{3}$ for $F=1$ atoms. For $F=1$ Rb atoms, $a_0\approx 101.8a_B$ and $a_2\approx 100.4a_B$~\cite{SK13}, implying $c_1\approx 0$ and our approximation to Eq.~\eqref{eq-F1} is a reasonable starting point. Here $a_B$ is the Bohr radius. The residual $c_1<0$ implies the ferromagnetic phase with normalized $|\mathbf{F}|=1$ may be more stable~\cite{SK13}. Fig.~\ref{fig-Fz} shows typical behavior of $F_z$ as the angle $\theta$ varies for the lump-solitons from the analytic solution with $N=1$ and the numerical solutions with $N=-1$ and $N=-2$. One can see that the solutions have inhomogeneous ferromagnetic order compatible with $c_1<0$ for Rb atoms.

We mention that one type of topological excitations from the magnetization in the presence of finite $c_1$ and external magnetic field is the skyrmion~\cite{Kawaguchi12}, which has been realized in spinor BEC as well~\cite{PhysRevLett.108.035301}. The lump solitons on a sphere discussed here may be viewed as the counterpart of the skyrmions, as the former comes from the topology of the condensate wavefunctions with the quantized winding number while the latter comes from the topology of the magnetization with the quantized skyrmion number.

\section{Conclusion}\label{sec:Conclusion}
In conclusion, we have shown that spherical lump solitons are topological excitations that may be realizable in a $F=1$ spinor BEC on a spherical shell without any gauge field because the vector field of the condensate wavefunctions wraps the order-parameter space around the real space. The homotopy is characterized by the winding number of the lump solitons. In addition to the exact solutions from spherical harmonics, we have demonstrated an ansatz for generating lump soliton solutions of arbitrary winding numbers in principle. The lump solitons on sphere not only offer concrete examples of topological excitations of interacting quantum systems in curved space but also open a way for investigating nonlinear physics via geometry and topology.

\begin{acknowledgments}
Y. H. was supported by the NNSF of China (No. 11874272) and Science Specialty Program of Sichuan University (No. 2020SCUNL210). C. C. C. was supported by the NSF (No. PHY-2011360).
\end{acknowledgments}

\appendix



\section{Lump solitons on a plane}
For the planar case with $x_1=x$, $x_2=y$, and the metric with $h_1=h_2=1$, Eq.~(\ref{eq-CR}) in the $c_0\rightarrow\infty$ limit reduces to
\be
(\p_1+i\p_2)R=2\p_{\bar{z}}R=0.
\ee
Here we introduce $z=x+iy$ and $\bar{z}=x-iy$, then $\p_{\bar{z}}=\frac12(\p_1+i\p_2)$. The above equation is the Cauchy-Riemann equation, which means $R$ is a holomorphic function depending only on $z$ but not $\bar{z}$. Thus, the lump solution can be taken as a rational holomorphic function of the form
\be
R(z)=\frac{p(z)}{q(z)},
\ee
where $p(z)$ and $q(z)$ are polynomials of $z$ with no common factors. In this case, one can verify that the winding number is
\be
N=\textrm{max}\Big[\text{deg}(p),\,\text{deg}(q)\Big].
\ee
Here deg$(p)$ means the order of the polynomial $p(z)$.


\bibliographystyle{apsrev}

\end{document}